\documentclass{elsart}
\usepackage{graphicx}
\begin{document}
\begin{frontmatter}
\title{Magnetic fluctuations and superconductivity in YbPd$_2$Sn}
\author[psi1]{A. Amato},
\author[psi2]{B.~Roessli}, 
\author[psi2]{P.~Fischer},
\author[cea]{N.~Bernhoeft}, 
\author[ill]{A.~Stunault},
\author[psi1]{C.~Baines}, 
\author[japan]{A.~ D\"onni} and
\author[japan_2]{H.~Sugawara}
\address[psi1]{Lab. for Muon-Spin Spectroscopy, Paul Scherrer Institute, CH-5232 Villigen
     PSI, Switzerland}
\address[psi2]{Lab. for Neutron Scattering, Paul Scherrer Institute \& ETH
Zurich, CH-5232 Villigen
     PSI, Switzerland}
\address[cea]{CEA-Grenoble, Av. des Martyrs, F-38054 Grenoble cedex, France}
\address[ill]{Institut Laue-Langevin,  Av. des Martyrs, F-38054 Grenoble cedex, France}
\address[japan]{Dept. of Physics, Niigata University, Ikarashi 2-8050, Niigata 950-2181, Japan}
\address[japan_2]{Dept. of Physics, Tokyo Metropolitan Univ.,
Minami-Ohsawa 1-1, Hachioji-Shi, Tokyo 192-0397, Japan}
\begin{abstract}
We report muon spin relaxation and inelastic neutron measurements on the Heusler system 
YbPd$_2$Sn. Localised anisotropic and quasi-elastic Yb magnetic fluctuations are observed 
below $T = 150$K. Both $\mu$SR and neutron data indicate a slowing-down of the 
spin-fluctuations, upon lowering the temperature, similar to that observed in 
Kondo lattices. The temperature dependence of the quasi-elastic neutron signal is 
compatible with a crystal-electric field scheme having a $\Gamma_7^{\rm CEF}$ ground state. The 
muon depolarization rate exhibits an additional contribution upon decreasing 
the temperature below $T_{\rm c}$ suggesting a close interplay between magnetic fluctuations 
and the superconducting state.  
\end{abstract}
\end{frontmatter}
\underline{Corresponding author:} \\
Alex Amato,
Paul Scherrer Institute, CH-5232 Villigen PSI, Switzerland\\
Phone : +41-56-310 32 32; Fax : +41-56-310 32 94; Email : alex.amato@psi.ch
\newpage

The interplay between magnetism and superconductivity (SC) has been attracting 
interest in view of the possible r\^ole played by the magnetic fluctuations 
in the pairing mechanism between electrons leading, for example, 
to the formation of Cooper pairs 
with non-zero orbital moment \cite{nature}. 
In this context much effort is presently devoted either to systems which are 
close to a quantum critical point, i.e. where the superconducting state appears to be 
favored by the presence of critical spin-fluctuations, or to the so-called heavy-fermion 
compounds where the strong hybridization between the $f$ and conduction electrons 
appears to be the necessary ingredient for the occurrence of unconventional SC. 
On the other hand, rare-earth based systems like RRh$_4$B$_4$, Chevrel phases and 
RNi$_2$B$_2$C (where R represents the rare-earth ion) may be modelled in terms of spatially 
isolated electron subsystems, where the 4$f$ magnetic moments interact only weakly 
with the superconducting electrons. In this light, the Heusler system YbPd$_2$Sn plays a 
particular r\^ole, since on one side the proportion of magnetic $f$-ions is high and 
on the other side the nearest-neighbor distance between rare-earth ions is 
rather short ($\sim$4.5\AA) preventing a clear isolation between the magnetic and 
superconducting sublattices \cite{ishikawa,kierstead}.
 
YbPd$_2$Sn exhibits coexistence of SC ($T_{\rm c} = 2.3$K) and simple antiferromagnetism 
($T_{\rm N} = 0.28$K) characterized by a propagation vector {\bf k}$ = (001)$ and magnetic 
moments aligned along the [111] direction \cite{donni}.
The value of the ordered magnetic Yb moments (1.4$\mu_{\rm B}$) is slightly smaller than the value 
1.71$\mu_{\rm B}$ expected from the crystal-field ground state doublet $\Gamma_7^{\rm CEF}$. 
Starting from the parent non-magnetic and superconducting system YPd$_2$Sn ($T_{\rm c} = 4.55$K), 
it was shown that the depression of $T_{\rm c}$ in YbPd$_2$Sn is characterized 
by an anomalously large value of the product $N(0)J^2_{\rm sf}$, where $N(0)$ 
is the density of states at the Fermi level per atom per spin direction 
and $J_{\rm sf}$ represents the exchange interaction between the conduction electron spins and 
the localized Yb $4f$ spin \cite{malik}. This points towards a strong 
hybridization and Kondo-type resonant scattering between localized 
and conduction electrons in this system.
 
In this light, to gain insight into their possible r\^ole on the superconducting state, 
it is of interest to characterize the thermal evolution of the 
spin-fluctuation spectrum at low temperatures. In this paper 
we present muon-spin relaxation and inelastic neutron scattering (INS) measurements performed 
at the Paul Scherrer Institute (surface muons instruments GPS and LTF; cold-neutron 
three-axis spectrometer TASP) and at the Institut Laue-Langevin (cold-neutron three-axis 
spectrometer IN14). For all measurements, we used polycrystalline samples from the same batch 
which was previously investigated at low temperature by neutron diffraction~\cite{donni}.

The $\mu$SR measurements were performed with conventional zero-field (ZF) and 
longitudinal-field (LF) techniques \cite{amato} using either a continuous 
$^4$He-cryostat (GPS, base temperature 1.7K) or a dilution $^3$He-$^4$He refrigerator 
(LTF, base temperature $\sim 30$mK). For both instruments the same sample was investigated 
(powder glued on a high purity sample silver holder by mixing it with less than 2\% GE varnish). 
The neutron measurements were made at fixed final wavevector with the sample 
enclosed in an Al-container in a 
$^4$He cryostat of ILL-type which attains the base temperature of $T = 1.5$K. 
In order to characterize the inelastic signal, the measurements had to be performed 
with different configurations yielding energy resolutions at elastic positions 
between $\sim 50 \mu$eV and $\sim 0.8$ meV. In all cases a filter (cold Be or 
pyrolitic graphite) was installed in the beam to avoid contamination of the signal 
by higher-order wavelengths.

\begin{figure}
  \includegraphics[scale=1]{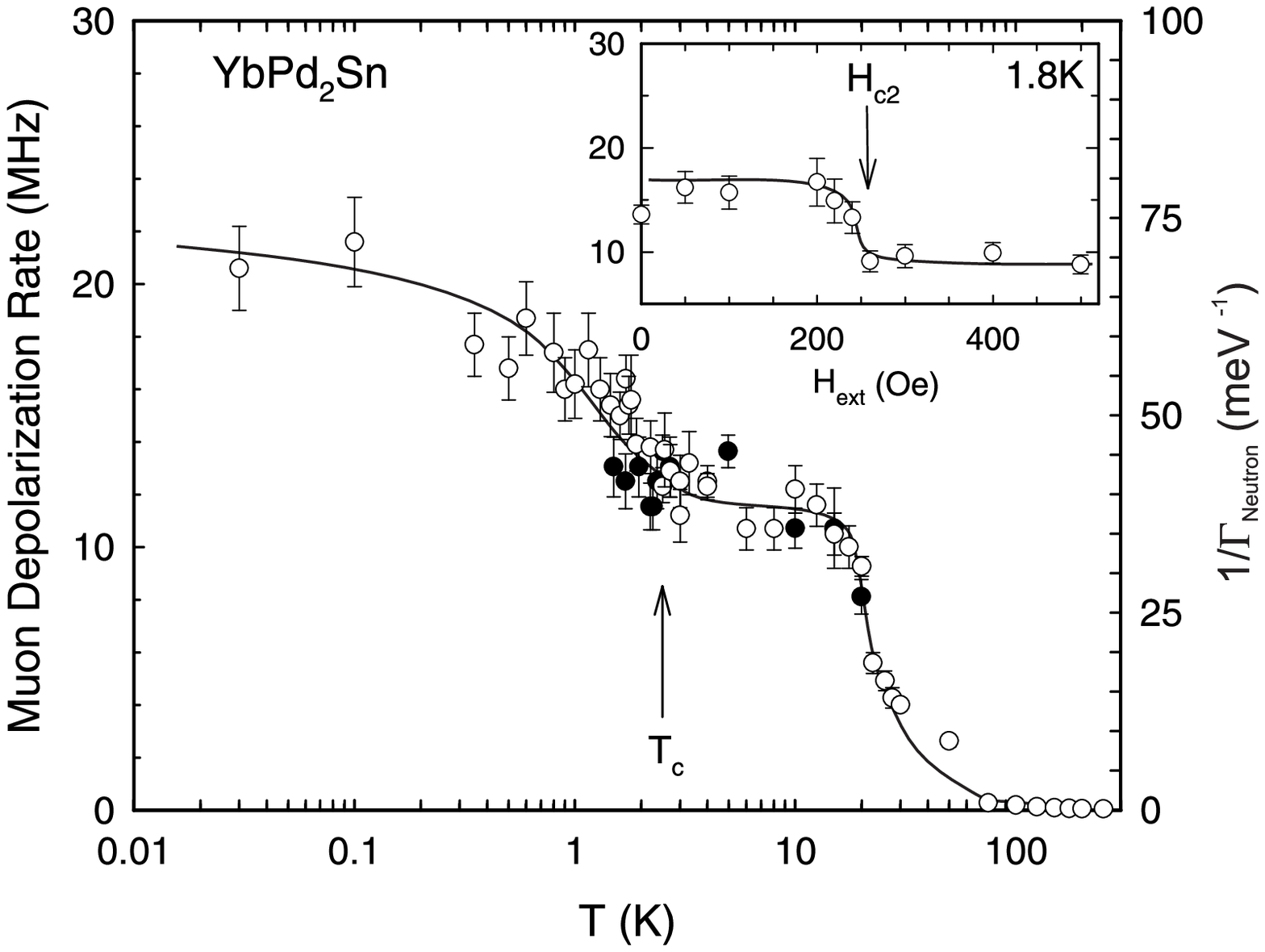}
   \caption{Temperature dependence of the fast component of the ZF muon 
    depolarization rate (open symbols, left axis). To allow comparison, the inverse of the neutron 
    quasi-elastic linewidth is also reported (closed symbols, right axis). Both quantities are 
    related through Eqs. \ref{fast}, \ref{cs} and \ref{tau}. The inset shows the field 
    dependence of the muon depolarization rate measured in LF at 1.8K where 
    the data were obtained in a field-cooling procedure.} 
   \label{musr}
\end{figure}

Whereas for temperatures above 50K, the ZF $\mu$SR signal is described by a single 
exponential function, at lower temperature it is best fitted assuming the depolarization function
\begin{equation}
P_{\mu}(t) = A_{slow}\exp(-\lambda_{slow} t) + A_{fast}\exp(-\lambda_{fast} t).
\label{depol_function}
\end{equation}
Below $T = 50$K, both the ratio between the  
amplitudes and the depolarization rate $\lambda_{slow}$ are found 
to be temperature independent with $A_{slow}/A_{fast}= 0.50(5)$ and 
$\lambda_{slow} \sim 0.55(3)$ MHz, whilst $\lambda_{fast}$ varies 
as shown in Fig. \ref{musr}. 
The exponential character of the muon polarization decay, 
as well as the observation of a similar signal in LF measurements 
($H_{\rm ext} = 2$kOe) performed above $T_{\rm c}$, reveal the existence of fast 
Yb spin fluctuations measurable within the $\mu$SR time-window. 
In addition, the presence of a two-component depolarization function of 
the observed amplitude ratio is indicative of a strong spatial anisotropy 
of such spin fluctuations producing an anisotropic field distribution at 
the muon site (i.e. $M_{2,\parallel} \gg M_{2,\perp}$ 
where $M_{2,i}$ represents the second moment of the static field 
distribution at the muon site along different directions). 
The portion of the muon ensemble possessing an initial polarization along 
the principal direction of the field distribution at the muon site is reflected 
by the first component of
Eq.~\ref{depol_function}, with 
\begin{equation}
\lambda_{slow} = \tau \gamma^2_{\mu} 2M_{2,\perp}~,
\end{equation}
where $\gamma_{\mu}$ is the gyro-magnetic ratio of the muon and $\tau$ represents 
the characteristic time of the anisotropic Yb-spin fluctuations. The second 
component of Eq.~\ref{depol_function} is dominated by the other spin directions
and its depolarization rate can be expressed as 
\begin{equation}
\lambda_{fast} = \tau \gamma^2_{\mu} (M_{2,\parallel} +M_{2,\perp}) \simeq \tau \gamma^2_{\mu} M_{2,\parallel}~.
\label{fast}
\end{equation}
Additional information on the spin-fluctuations is obtained from inelastic neutron 
measurements for which the cross section reflects the imaginary part of the wave-vector 
and frequency dependent susceptibility $\chi^{''}(\mbox{\boldmath $\rm q$},\omega)$. 
In a single ion approximation, crystal electric field (CEF) transitions yield a
{\bf q}-independent susceptibility: 
\begin{equation}
{\chi^{''} (\omega)\over{\pi \omega}}=\sum_{m}\chi^m_{\rm C} P_{mm}(\omega)
+{1\over2}\sum_{m \neq n}\chi^{nm}_{\rm vv}[1-\exp(-\beta \Delta_{nm})]P_{nm}(\omega
-\Delta_{nm})
\label{cs}
\end{equation}
where the sums go over the CEF-levels and $\chi_{\rm C}$ and $\chi_{\rm vv}$ are the Curie and 
Van Vleck susceptibilities, respectively \cite{dalmas}.
When dominated by single relaxation rate decay of the magnetisation, the scattering function
$P_{nm}(\omega)$ reads 
\begin{equation}
P_{nm}(\omega)={1 \over \pi}{{\Gamma_{nm} (T)}\over {\omega^2+\Gamma^2_{nm}} (T)}.
\label{lor}
\end{equation}   
High-resolution inelastic neutron measurements in YbPd$_2$Sn have indeed revealed 
a quasi-elastic signal characteristic of localized excitations arising from single-site 
fluctuations with the energy dispersion of the form of Eq.~\ref{lor}. An effective CEF scheme 
can be inferred from the temperature dependence of the frequency integrated weight of the quasi-elastic signal. 
In Fig.~\ref{neutron_and_cef} the best fit to the experimental data is given. 
The fit yields the cubic CEF parameters $W=-6.629$K and $x=-0.740$ which are close 
from the values reported by Li \textit{et al.} \cite{li}. With these parameters, the excited $\Gamma_8^{\rm CEF}$ 
state lies 4meV above the $\Gamma_7^{\rm CEF}$ ground state (see inset Fig.~\ref{neutron_and_cef}) 
which is in reasonable agreement with our INS data which reveal a  
transition at 3.6meV (not shown). 

In the absence of a {\bf q}-dependence of the Yb-fluctuations, 
the  local probe $\mu$SR data can directly be connected to the dynamical susceptibility since
\begin{equation}
\tau = \frac{1}{\Gamma_{\mu{\rm SR}}} = \frac{1}{\chi_{\rm C}}\lim_{\omega\to 0}\frac{\chi^{''}(\omega)}{\pi\omega}.
\label{tau}
\end{equation}
For a single ion in the presence of Kondo coupling, the dynamical susceptibility 
has been obtained within the NCA approximation by Cox et al. \cite{cox}. 
At very low temperature (i.e. below the characteristic Kondo temperature), 
the dynamical susceptibility assumes a single Lorentzian form with a constant 
dynamical width, whereas a $T^{\beta}$-like temperature dependence of this latter 
with $\beta = 1/2$ is 
expected at high temperature when the full degeneracy of the Yb-spin configuration 
is recovered. 
The experimental thermal variation of $\Gamma_{\mu{\rm SR}} \propto \lambda_{fast}^{-1}$ 
and of the width of the quasi-elastic Lorentzian neutron signal do indeed 
show a constant value at low temperatures just above $T_{\rm c}$. 
On the other hand the temperature dependence at high temperatures suggests an exponent 
$\beta$ value much higher than 1/2. Such a behavior could reflect the presence of 
CEF effects and/or muon diffusion effects acting on the 
observed muon depolarization at high temperatures. 
If on one hand the importance of CEF effects is confirmed by the level 
scheme extracted from our INS 
measurements (see Fig.~\ref{neutron_and_cef}), on the other hand the presence of 
muon diffusion at high temperatures, 
which could mask the  effects of the Yb-fluctuations on $\lambda_{fast}$,
is suggested by the observation of a simple exponential depolarization. 
Assuming muons not to diffuse at high temperatures, an initial Gaussian 
depolarization should have been observed, 
reflecting the expected Gaussian field distribution at the muon site 
caused by the nuclear moments 
of the $^{171,173}$Yb, $^{105}$Pd and $^{115,117,118}$Sn isotopes \cite{schenck}. 
The observed exponential decay of the polarization demonstrates the mobility of 
the muon at high temperature 
leading to motional narrowing phenomena masking the intrinsic 
effects of the Yb-fluctuations 
on the field distribution at the muon site.

As shown in Fig.~\ref{musr}, the $\mu$SR depolarization rate $\lambda_{fast}$ 
exhibits a clear increase upon decreasing
the temperature below $T_{\rm c}$. The relation between the observed increase 
and the presence of the superconducting 
state is confirmed by measurements performed in the LF configuration 
for temperatures below $T_{\rm c}$ as shown in the
inset of Fig. ~\ref{musr}. These measurements performed at 
1.8K in field-cooling procedures indicate that the additional
increase of the depolarization rate occurring below $T_{\rm c}$ is canceled when the external magnetic field is higher
than the critical field $H_{c2}(T=1.8{\rm K})\simeq 250$G \cite{aoki}. 

In the following we discuss different possible
origins for the increase of the $\mu$SR depolarization rate in the superconducting state. An obvious origin is related
to the formation of the superconducting gap which decreases the number of available conduction electrons capable to
participate to a Kondo interaction with the localized 4$f$-spin. However, the reduction of the ordered magnetic
moment, below $T_{\rm N}$ determined by neutron scattering experiments \cite{donni}, compared to the calculated value
expected from the $\Gamma_7^{\rm CEF}$ ground state, suggests that the Kondo interaction is still effective much below
$T_{\rm c}$. Another possibility is the presence of short-range correlations suggested by specific heat measurements
where an anomalous broadening of the Schottky peak caused by the Zeeman splitting of the $\Gamma_7^{\rm CEF}$ ground state
is observed \cite{aoki}. Such short-range dynamical correlations were found to coexist with single-site (i.e.
{\bf q}-independent) fluctuations in a number of Kondo systems \cite{rossat}, but the weight of their signals
integrated in $q$-space is usually much smaller than the weight of the quasi-elastic contribution. This fact 
could explain the absence of an anomalous behavior of the neutron line-width measured at a given {\bf q}-value. 
Interestingly, the specific heat data indicate that the short-range correlations are detectable only below the 
superconducting transition, as also suggested by the $\mu$SR data, pointing to a possible interplay between the 
occurrence of the superconducting state and the magnetic correlations. 

We thank Prof. H. Sato and Dr. Y. Aoki for fruitful discussions and the help during the sample preparation. This work was partly supported by the European Science Foundation (FERLIN programme).

\newpage

\begin{figure}
   \includegraphics[scale=1]{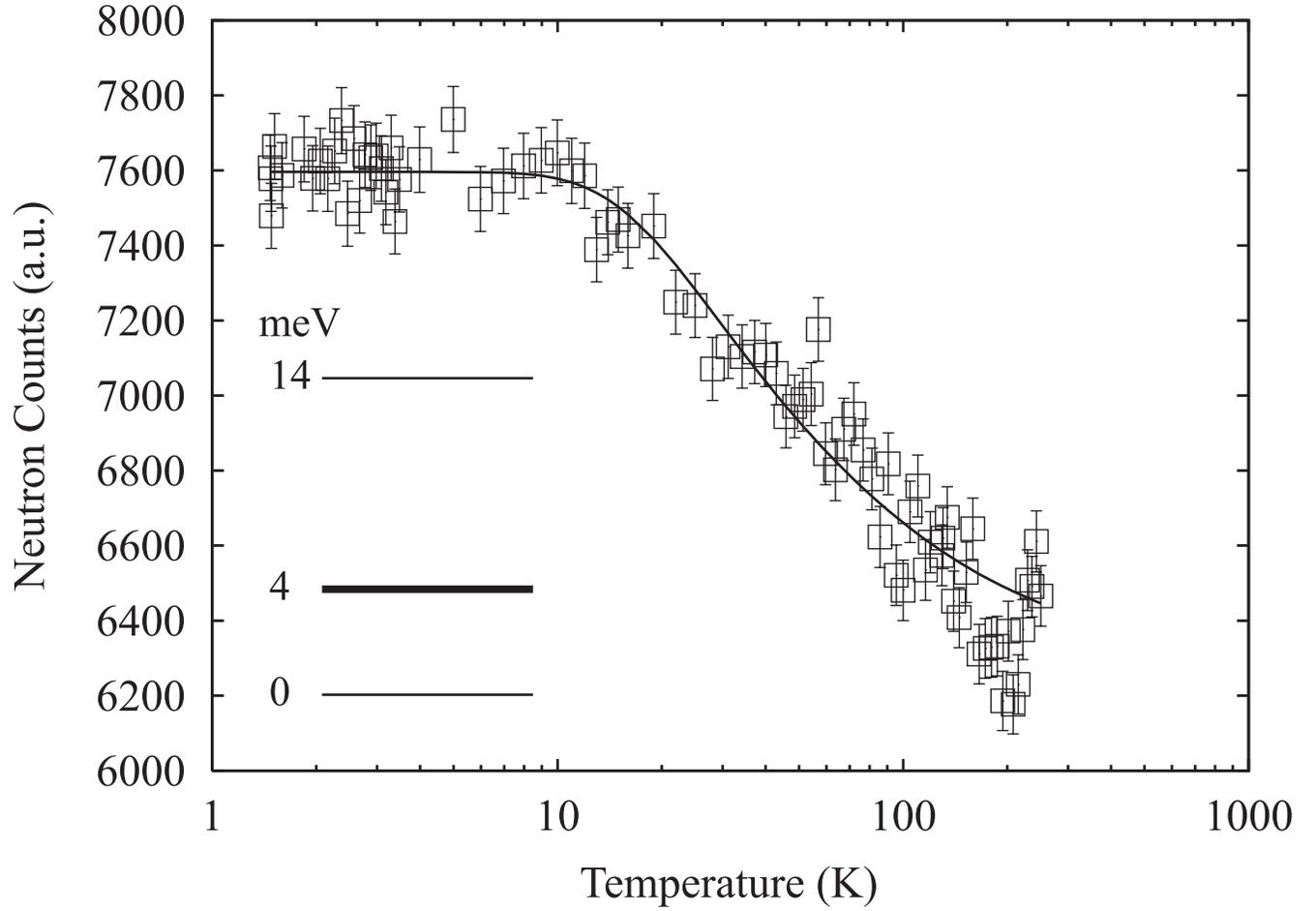}
   \caption{\label{neutron_and_cef} Temperature dependence of the integrated weight 
   of the quasi-elastic signal obtained at ${\rm Q} = 0.15$~ \AA$^{-1}$. The line represents the best fit obtained with Eq.~\ref{cs}.
   From the fit parameters, the CEF shown in the inset is calculated (ground state 
   $\Gamma_7^{\rm CEF}$ doublet; first excited state $\Gamma_8^{\rm CEF}$ quadruplet; and second excited state 
   $\Gamma_6^{\rm CEF}$ doublet).} 
   \label{neutron_and_cef}
\end{figure}
\end{document}